\documentclass[dvips]{acta}
\usepackage{supertabular,lscape,epsfig}
\usepackage{amssymb}
\usepackage{amsmath}
\usepackage{float}
\mathchardef\mhyphen="2D
\usepackage{placeins}
\DeclareSymbolFont{ppa}{OT1}{ppl}{m}{it}
\DeclareMathSymbol{\vv}{\mathalpha}{ppa}{'166}

\SetPages{0}{0}

\SetVol{67}{2017}

\begin{document}
\newcommand\pvalue{\mathop{p\mhyphen {\rm value}}}
\newcommand{\TabApp}[2]{\begin{center}\parbox[t]{#1}{\centerline{
  {\bf Appendix}}
  \vskip2mm
  \centerline{\small {\spaceskip 2pt plus 1pt minus 1pt T a b l e}
  \refstepcounter{table}\thetable}
  \vskip2mm
  \centerline{\footnotesize #2}}
  \vskip3mm
\end{center}}

\newcommand{\TabCapp}[2]{\begin{center}\parbox[t]{#1}{\centerline{
  \small {\spaceskip 2pt plus 1pt minus 1pt T a b l e}
  \refstepcounter{table}\thetable}
  \vskip2mm
  \centerline{\footnotesize #2}}
  \vskip3mm
\end{center}}

\newcommand{\TTabCap}[3]{\begin{center}\parbox[t]{#1}{\centerline{
  \small {\spaceskip 2pt plus 1pt minus 1pt T a b l e}
  \refstepcounter{table}\thetable}
  \vskip2mm
  \centerline{\footnotesize #2}
  \centerline{\footnotesize #3}}
  \vskip1mm
\end{center}}

\newcommand{\MakeTableApp}[4]{\begin{table}[p]\TabApp{#2}{#3}
  \begin{center} \TableFont \begin{tabular}{#1} #4 
  \end{tabular}\end{center}\end{table}}

\newcommand{\MakeTableSepp}[4]{\begin{table}[p]\TabCapp{#2}{#3}
  \begin{center} \TableFont \begin{tabular}{#1} #4 
  \end{tabular}\end{center}\end{table}}

\newcommand{\MakeTableee}[4]{\begin{table}[htb]\TabCapp{#2}{#3}
  \begin{center} \TableFont \begin{tabular}{#1} #4
  \end{tabular}\end{center}\end{table}}

\newcommand{\MakeTablee}[5]{\begin{table}[htb]\TTabCap{#2}{#3}{#4}
  \begin{center} \TableFont \begin{tabular}{#1} #5 
  \end{tabular}\end{center}\end{table}}

\newcommand{\MakeTableH}[4]{\begin{table}[H]\TabCap{#2}{#3}
  \begin{center} \TableFont \begin{tabular}{#1} #4 
  \end{tabular}\end{center}\end{table}}

\newcommand{\MakeTableHH}[4]{\begin{table}[H]\TabCapp{#2}{#3}
  \begin{center} \TableFont \begin{tabular}{#1} #4 
  \end{tabular}\end{center}\end{table}}

\newfont{\bb}{ptmbi8t at 12pt}
\newfont{\bbb}{cmbxti10}
\newfont{\bbbb}{cmbxti10 at 9pt}
\newcommand{\uprule}{\rule{0pt}{2.5ex}}
\newcommand{\douprule}{\rule[-2ex]{0pt}{4.5ex}}
\newcommand{\dorule}{\rule[-2ex]{0pt}{2ex}}
\begin{Titlepage}
\Title{Concluding Henrietta Leavitt's Work on Classical Cepheids\\
in the Magellanic System\\
and Other Updates of the OGLE Collection of Variable Stars\footnote{Based on observations
obtained with the 1.3-m Warsaw telescope at the Las Campanas Observatory of
the Carnegie Institution for Science.}}
\Author{I.~~S~o~s~z~y~ñ~s~k~i$^1$,~~
A.~~U~d~a~l~s~k~i$^1$,~~
M.\,K.~~S~z~y~m~a~ñ~s~k~i$^1$,~~
\L.~~W~y~r~z~y~k~o~w~s~k~i$^1$,\\
K.~~U~l~a~c~z~y~k$^2$,~~
R.~~P~o~l~e~s~k~i$^{1,3}$,~~
P.~~P~i~e~t~r~u~k~o~w~i~c~z$^1$,~~
S.~~K~o~z~³~o~w~s~k~i$^1$,\\
D.\,M.~~S~k~o~w~r~o~n$^1$,~~
J.~~S~k~o~w~r~o~n$^1$,~~
P.~~M~r~ó~z$^1$,~~
and~~M.~~P~a~w~l~a~k$^1$}
{$^1$Warsaw University Observatory, Al.~Ujazdowskie~4, 00-478~Warszawa, Poland\\
e-mail: soszynsk@astrouw.edu.pl\\
$^2$Department of Physics, University of Warwick, Gibbet Hill Road, Coventry, CV4~7AL,~UK\\
$^3$Department of Astronomy, Ohio State University, 140 W. 18th Ave., Columbus, OH~43210, USA}
\Received{June 23, 2017}
\end{Titlepage}

\Abstract{More than a century ago, Henrietta Leavitt discovered the
  first Cepheids in the Magellanic Clouds together with the famous
  period--luminosity relationship revealed by these stars, which soon after
  revolutionized our view of the Universe. Over the years, the number of
  known Cepheids in these galaxies has steadily increased with the
  breakthrough in the last two decades thanks to the new generation of
  large-scale long-term sky variability surveys.

  Here we present the final upgrade of the OGLE Collection of Cepheids in
  the Magellanic System which already contained the vast majority of known
  Cepheids. The updated collection now comprises 9649 classical and 262
  anomalous Cepheids. Type-II Cepheids will be updated shortly. Thanks to
  high completeness of the OGLE survey the sample of classical Cepheids
  includes virtually all stars of this type in the Magellanic Clouds.
  Thus, the OGLE survey concludes the work started by Leavitt.

  Additionally, the OGLE sample of RR~Lyr stars in the Magellanic System
  has been updated. It now counts 46\,443 variables. A collection of seven
  anomalous Cepheids in the halo of our Galaxy detected in front of the
  Magellanic Clouds is also presented.

  OGLE photometric data are available to the astronomical community from
  the OGLE Internet Archive. The time-series photometry of all pulsating
  stars in the OGLE Collection has been supplemented with new
  observations.}{Stars: variables: Cepheids -- Stars: variables: RR~Lyrae
  -- Magellanic Clouds -- Catalogs}

\Section{Introduction}
\vspace*{5pt} 
Since the discovery of the first Cepheids in the Magellanic Clouds by
Henrietta Leavitt (1908) and their famous period--luminosity relation (Leavitt and
Pickering 1912), the catalogs of variable stars in these galaxies have been
steadily growing. Up to the beginning of 1990s, efforts of many observers
led to the discovery of about 2500 Cepheids and 300 RR~Lyr stars in both
Clouds (Artyukhina \etal 1995 and references therein). Then, large-scale
optical sky surveys, in particular the MACHO (MAssive Compact Halo Object,
Alcock \etal 1995) and OGLE (Optical Gravitational Lensing Experiment,
Udalski \etal 2015) projects, increased by a large factor the number of
known variable stars in the Magellanic System. The most recent edition of
the OGLE Collection of Variable Stars (OCVS) contains 250 anomalous
Cepheids (Soszyñski \etal 2015a), 9535 classical Cepheids (Soszyñski \etal
2015b), and 45\,451 RR~Lyr stars (Soszyñski \etal 2016a) detected in about
650 square degrees covering the Large Magellanic Cloud (LMC), Small
Magellanic Cloud (SMC), and Magellanic Bridge.

These OGLE samples of pulsating stars have already been used in many
studies. For example, for the investigation of the three-dimensional
Magellanic System structure (Jacyszyn-Dobrzeniecka \etal 2016, 2017, Inno
\etal 2016), with particular emphasis on the Magellanic Bridge connecting
both Magellanic Clouds (Belokurov \etal 2017, Wagner-Kaiser and Sarajedini
2017), for constructing metallicity maps of the Magellanic Clouds (Skowron
\etal 2016), comparing observational and theoretical light curves of
Cepheids (Bhardwaj \etal 2017, Marconi \etal 2017), or detecting peculiar
forms of double-periodic pulsations in Cepheids and RR~Lyr stars (Smolec
\etal 2016, Smolec and ¦niegowska 2016, Smolec 2017, Soszyñski \etal
2016b).

The OGLE collections of pulsating stars published so far are practically
complete in the central regions of the Magellanic Clouds (about 40 square
degrees in the LMC and 14 square degrees in the SMC), because these fields
were monitored during the previous phases of the OGLE survey (OGLE-II and
OGLE-III) and were independently searched for variable stars. However,
external regions of both galaxies have only been observed during the
ongoing fourth phase -- OGLE-IV. The completeness of the OGLE samples of
variable stars identified in these areas has been reduced by technical gaps
between CCD detectors of the OGLE mosaic camera. About 7\% of stars fell
into these gaps which decreased the completeness of our collection by the
same factor.

In this paper we supplement the OGLE Collection of Cepheids and RR~Lyr
stars in the Magellanic Clouds with objects detected in these regions. The
updated collection now contains practically all classical Cepheids. Thus,
with the update presented here we conclude the work on classical Cepheids
started by Leavitt. The upgraded OCVS also contains the vast majority
of RR~Lyr stars and anomalous Cepheids in the Magellanic System covered by
the OGLE-IV fields.

\newpage
The paper is organized as follows. Section~2 presents the dataset used in
this investigation. In Section~3 we briefly describe how pulsating stars
were selected. The upgraded collection of Cepheids and RR~Lyr stars in the
Magellanic Clouds is presented in Section~4, while a sample of anomalous
Cepheids belonging to our Galaxy is discussed in Section~5. Section~6 is
dedicated to the completeness estimation and comparison with the results
from other large-scale sky surveys. The paper is concluded in Section~7.

\Section{Observations and Data Reduction}
The observational data used in this study come from the auxiliary
photometric databases of the OGLE-IV project and were obtained from 2010 to
2016 with the 1.3-meter Warsaw Telescope at Las Campanas Observatory,
Chile. The observatory is operated by the Carnegie Institution for
Science. The Warsaw Telescope is equipped with a 256 Megapixel mosaic
camera composed of 32 CCD detectors (Udalski \etal 2015). The standard high
precision OGLE photometry of the Magellanic System fields was obtained with
the OGLE data pipeline (Udalski 2003, Udalski \etal 2015) based on the
Difference Image Analysis (DIA) method (Alard and Lupton 1998, Wo¼niak
2000).

The OGLE-IV camera has several technical gaps between the detectors, which
results in strips of ``dead zones'' in the sky, \ie regions inaccessible
during a single exposure. The strips are from 17\arcs to 97\arcs wide which
decreases the completeness of the regular OGLE-IV databases by about
7\%. To minimize this problem, a dithering technique was applied to the
OGLE-IV observations during two observing seasons. Additionally, a natural
dithering caused by imperfections in the telescope pointing (${\it
rms}\approx15\arcs$) systematically filled the gaps between the CCD
detectors. One should, however, remember that the typical number of
observations secured in these regions is significantly smaller than in the
remaining parts of the fields.

The OGLE-IV collection of images is so large that it is now possible to
create deep, good-seeing reference images filling practically the entire
area covered by the mosaic camera. Therefore we decided to repeat the DIA
reductions focusing on the regions that were not covered by the standard
OGLE-IV reductions before.

The new DIA reference images for each OGLE-IV field in the Magellanic
System were constructed from 50--100 best individual {\it I}-band images,
which resulted in much deeper photometry than in the regular reference
images composed typically of up to ten images. Our deep reference images
filled practically all ``dead zones''. The photometry of stars in these
regions was obtained using the standard OGLE pipeline (Udalski 2003)
running with this new set of deep reference images. We extracted photometry
for all objects detected in the ``dead zones'' and their surroundings (for
testing purposes). Additionally, we obtained time-series photometry of all
bright stars ($I<15$~mag) regardless of their position on the
detector. Some of these objects could have been saturated on the regular
OGLE reference images (which typically had better seeing) so their
photometry in the standard databases was not available. Now, with the new
reductions some of them could be measured.

The new auxiliary {\it I}-band OGLE databases of the Magellanic System
fields were created based on the new reductions. Because the new deep
reference images share the same flux scale as the standard ones, the
auxiliary databases are in the same photometric system as the standard OGLE
databases. The calibrations to the standard system were then carried out in
the identical way as for the standard databases (Udalski \etal 2015).

The total area covered by the extended OGLE-IV fields in the Magellanic
System reaches 670~square degrees. The number of data points in the new
auxiliary photometric database depends mainly on the position of a star in
the field. For the sample of the newly detected Cepheids and RR~Lyr stars
the median number of the {\it I}-band observations is 70, but for
individual stars we obtained from 13 to more than 600 data points.

Because the new reductions were carried out only for the {\it I}-band
images (for variability search), the {\it V}-band light curves come from
the standard DIA reductions, and about half of the newly detected variables
do not have the {\it V}-band measurements. For the remaining stars, the
median number of the {\it V}-band data points is seven, but in some cases
it exceeds 100.

\vspace*{-9pt}
\Section{Selection of Cepheids and RR~Lyr Stars}
\vspace*{-5pt}
The new auxiliary OGLE databases were searched for variable stars. The
procedure was in principle the same as that described by Soszyñski \etal
(2015a, 2016a), but we took special care to ensure the high completeness of
the variability detection in poorly sampled light curves. All stars with
more than ten {\it I}-band data points were searched for periodicity using
two different methods: the discrete Fourier transform implemented in the
{\sc Fnpeaks}
code\footnote{http://helas.astro.uni.wroc.pl/deliverables.php?lang=en\&active=fnpeaks}
and the Analysis of Variance algorithm. Our procedures were applied to
about 18~million light curves.

From this sample we visually inspected the light curves with the highest
signal-to-noise ratio and periods from 0.2~d to 50~d. We paid special
attention to stars located in the Cepheid instability strip in the
color--magnitude diagram ($0.3<V-I<1.1$~mag), lying within a wide strip in
the period--luminosity diagram covering all types of Cepheids and RR~Lyr
stars, and with {\it I}-band amplitudes larger than 0.1~mag. Based on the
light curve morphology, we selected candidates for pulsating stars,
eclipsing binaries, and other variable stars. Then, an analysis of periods,
mean luminosities, colors, Fourier parameters of the light curves, and
period ratios (in multiperiodic variables) allowed us to divide our sample
of pulsating stars into classical, type~II, anomalous Cepheids, and RR~Lyr
stars. Finally, objects in each group were divided into subtypes, according
to their pulsation modes. For completeness, we supplemented our list by
several well-known long-period classical Cepheids that are too bright to be
monitored by OGLE.

\begin{figure}[p]
\centerline{\includegraphics[width=12.7cm]{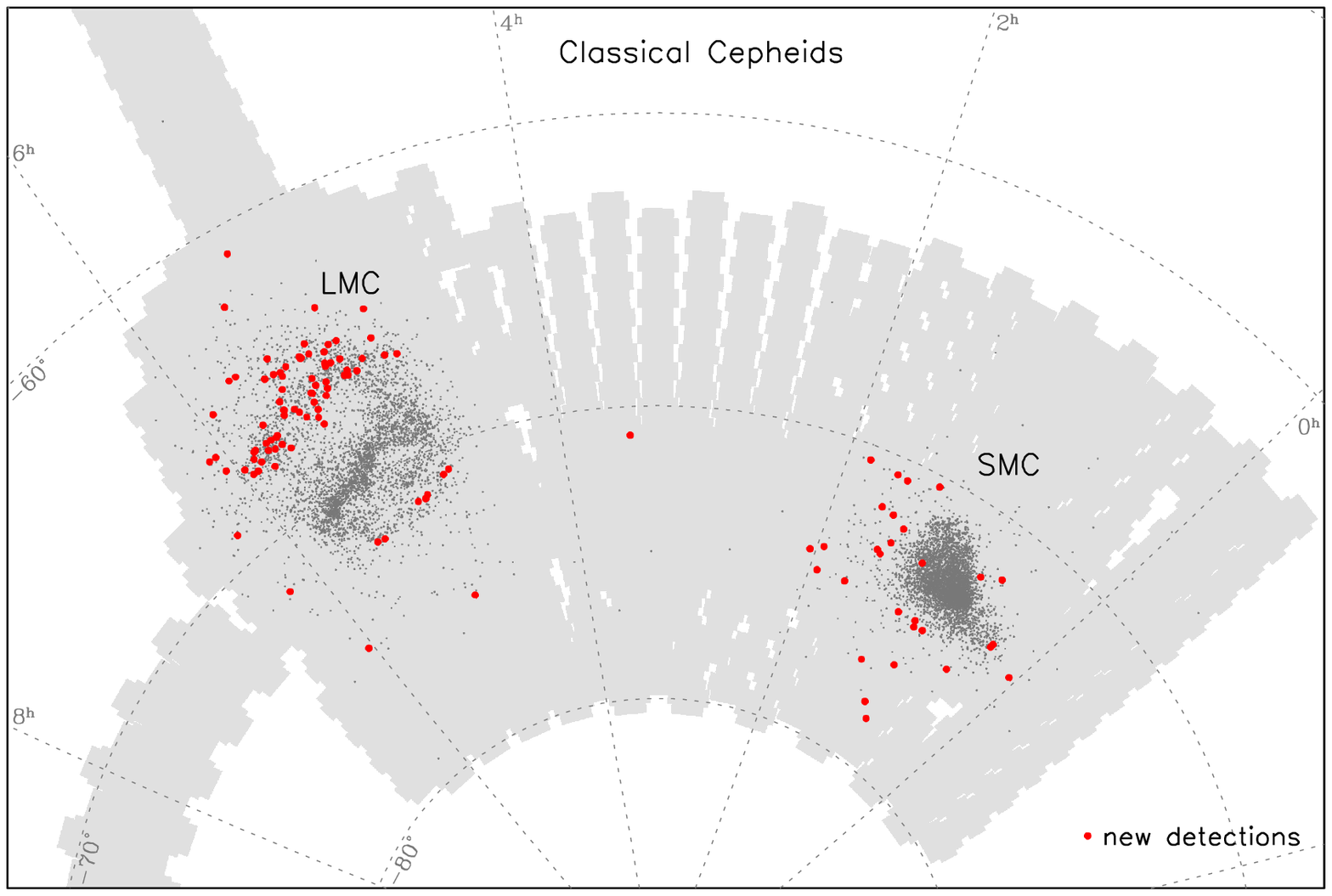}}
\vspace*{3pt}
\FigCap{Spatial distribution of classical Cepheids in the Magellanic
System. Dark gray points mark objects included in the previous edition
of the OCVS, while red points indicate newly detected Cepheids. The gray
area shows the sky coverage of the OGLE-IV fields.}
\vskip7mm
\centerline{\includegraphics[width=12.7cm]{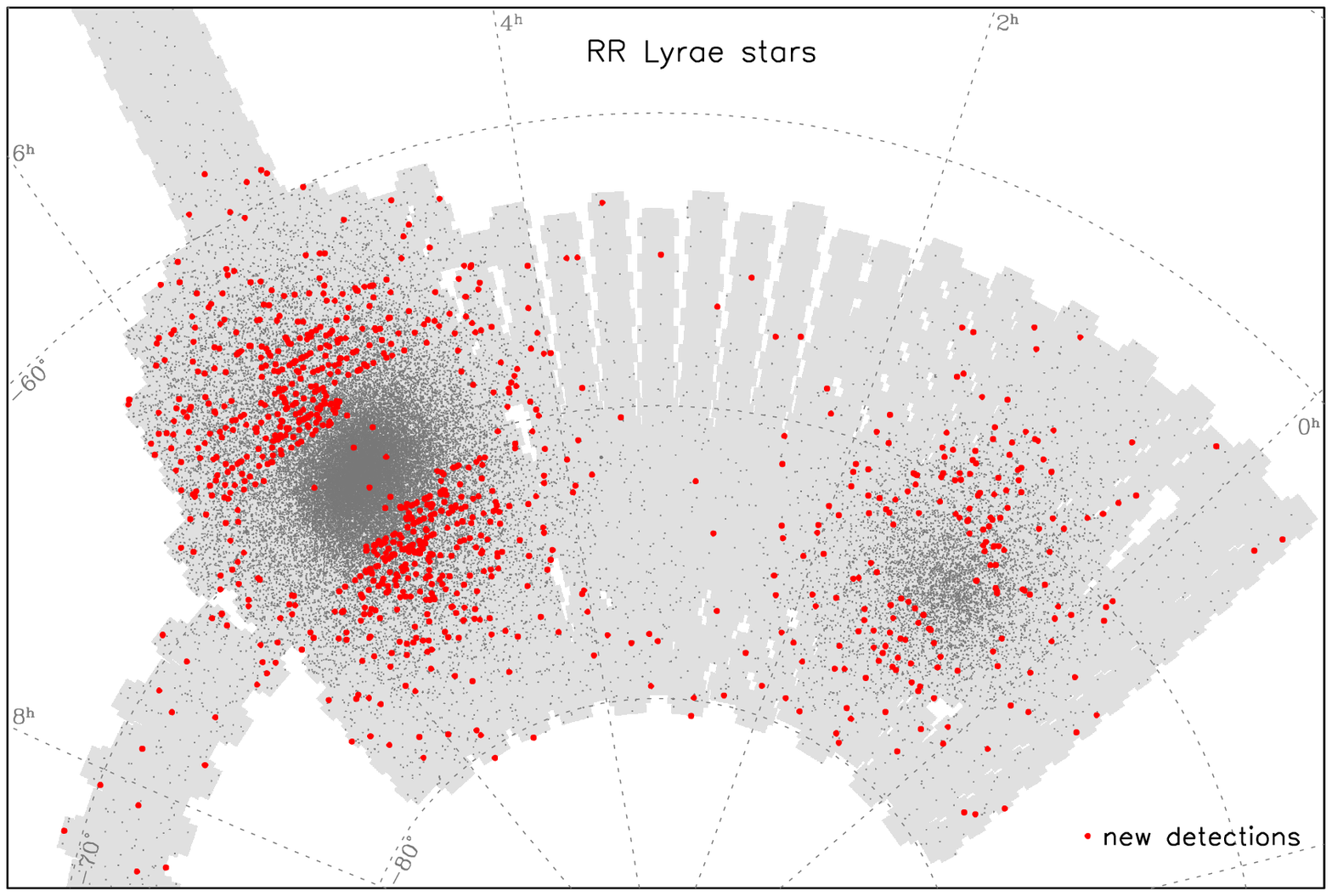}}
\vspace*{3pt}
\FigCap{Spatial distribution of RR~Lyr stars toward the Magellanic
System. Dark gray points mark variables included in the previous edition
of the OCVS, while red points indicate newly detected stars. The gray
area shows the OGLE-IV footprint.}
\end{figure}

Our search led to the identification of 115 classical Cepheids, 994 RR~Lyr
stars, and 12 anomalous Cepheids that were not included in the OCVS, so
far. Red points in Figs.~1--3 show the positions of these stars in the
sky. As expected, almost all new Cepheids and the vast majority of RR~Lyr
variables are located outside the OGLE-III fields. It is worth noting that
one of the newly detected classical Cepheids (OGLE-SMC-CEP-4987) lies in
the region of the Magellanic Bridge, a stream of neutral gas and stars that
links the two Clouds. It increases the sample of the Bridge Cepheids to
ten. Cepheids in the Magellanic Bridge were recently analyzed by
Jacyszyn-Dobrzeniecka \etal (2016).
\begin{figure}[htb]
\vglue7pt
\centerline{\includegraphics[width=12.7cm]{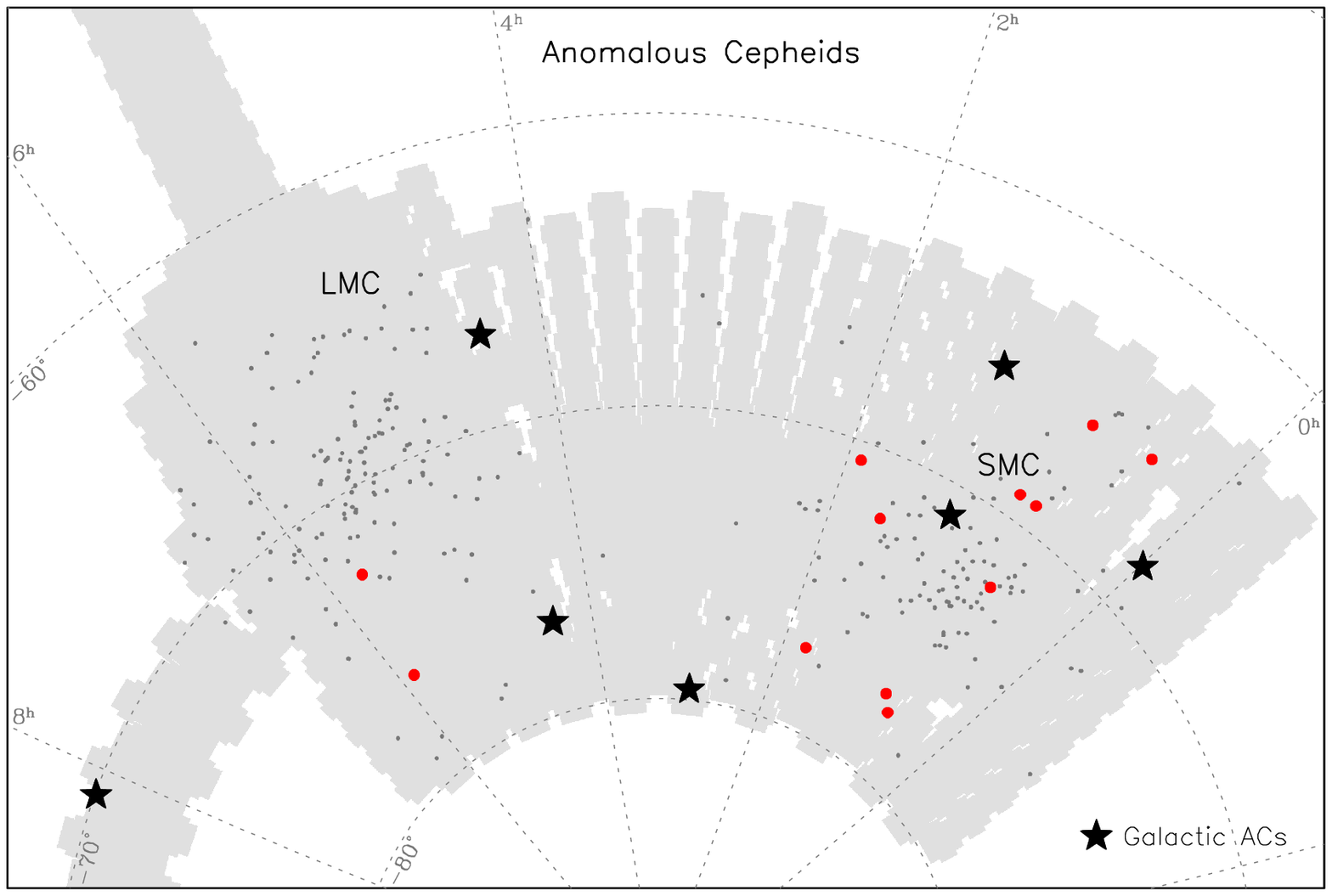}}
\vspace*{5pt}
\FigCap{Spatial distribution of anomalous Cepheids in the Magellanic
System. Dark gray points mark variables included in the previous edition of
the OCVS, red points indicate newly detected objects, and star symbols show
Galactic anomalous Cepheids in the foreground of the Magellanic Clouds. The
gray area shows the sky coverage of the OGLE-IV fields.}
\end{figure}

\Section{OGLE Collection of Cepheids and RR~Lyr Stars in the Magellanic Clouds}
The newly detected Cepheids and RR~Lyr stars have been added to the OCVS
(Soszyñski \etal 2015ab, 2016a). The full collection contains now 9\,649
classical Cepheids, 262 anomalous Cepheids, and 46\,443 RR~Lyr variables in
the Magellanic System. The exact numbers of stars pulsating in various
modes in both Clouds are summarized in Table~1. The collection of type II
Cepheids in the Magellanic Clouds will be published elsewhere.

\newpage
\MakeTableee{l@{\hspace{10pt}}
l@{\hspace{10pt}} 
r@{\hspace{10pt}} 
r@{\hspace{10pt}}
r@{\hspace{10pt}}}{12.5cm}{Numbers of Pulsating Stars in the Magellanic Clouds}
{\hline
\noalign{\vskip3pt}
\multicolumn{1}{c}{Type}
& Subtype
&\multicolumn{1}{c}{LMC}
&\multicolumn{1}{c}{SMC}
&\multicolumn{1}{c}{Total}\\
\noalign{\vskip3pt}
\hline
\noalign{\vskip3pt}
Classical Cepheids & all       &  $4\;704$ &  $4\;945$ &  $9\;649$ \\
                   & F         &  $2\;476$ &  $2\;753$ &  $5\;229$ \\
                   & 1O        &  $1\;775$ &  $1\;793$ &  $3\;568$ \\
                   & 2O        &      $26$ &      $91$ &     $117$ \\
                   & F/1O      &      $95$ &      $68$ &     $163$ \\
                   & 1O/2O     &     $322$ &     $239$ &     $561$ \\
                   & 1O/3O     &       $1$ &       $0$ &       $1$ \\
                   & 2O/3O     &       $1$ &       $0$ &       $1$ \\
                   & F/1O/2O   &       $1$ &       $0$ &       $1$ \\
                   & 1O/2O/3O  &       $7$ &       $1$ &       $8$ \\
\noalign{\vskip3pt}
\hline
\noalign{\vskip3pt}
Anomalous Cepheids & all       &     $143$ &     $119$ &     $262$ \\
                   & F         &     $102$ &      $78$ &     $180$ \\
                   & 1O        &      $41$ &      $41$ &      $82$ \\
\noalign{\vskip3pt}
\hline
\noalign{\vskip3pt}
RR~Lyrae Stars     & all       & $39\;871$ &  $6\;572$ & $46\;443$ \\
                   & RRab      & $28\;193$ &  $5\;105$ & $33\;298$ \\
                   & RRc       &  $9\;663$ &     $801$ & $10\;464$ \\
                   & RRd       &  $1\;995$ &     $663$ &  $2\;658$ \\
                   & anom. RRd &      $20$ &      $3$ &       $23$ \\
\noalign{\vskip3pt}
\hline}

The entire collection is available {\it via} anonymous FTP site:
\begin{center}
{\it ftp://ftp.astrouw.edu.pl/ogle/ogle4/OCVS/}\\
\end{center}

At this site, we provide basic parameters of each object: classification,
J2000 equatorial coordinates, pulsation periods (derived with the {\sc
Tatry} code by Schwar\-zenberg-Czerny 1996), {\it I}- and {\it V}-band
mean magnitudes, amplitudes, epochs of maximum light, Fourier coefficients,
and time series: {\it I}- and {\it V}-band photometry obtained from 2010 to
mid-2016 during the OGLE-IV project (earlier OGLE observations for some
stars are available in the OGLE-III Catalog of Variable Stars). The light
curves of pulsating stars released by Soszyñski \etal (2015ab, 2016a) have
been supplemented with new observations collected by the end of July
2016. Furthermore, we provide, for the first time, the OGLE-IV photometry
for about 2500 Cepheids and RR~Lyr variables known from the previous phases
of the OGLE survey (OGLE-II and OGLE-III), but located in the gaps between
the CCD chips of the OGLE-IV camera.

A few objects have been reclassified in this edition of the OCVS. A
short-period pulsator OGLE-LMC-CEP-1154 has been moved from the list of
classical Cepheids to the list of RR~Lyr stars (the regular OGLE light
curve of this star was affected by blending). Two stars previously
classified as RR~Lyr variables (OGLE-LMC-RRLYR-03358, OGLE-LMC-RRLYR-16235)
are likely eclipsing binaries. We removed these stars from the
collection. In several other RR~Lyr stars we corrected their pulsation
periods, which in one case (OGLE-LMC-RRLYR-21326) led to a modification of
the pulsating mode.

\Section{Galactic Anomalous Cepheids}
Until recently, only one anomalous Cepheid was known in the field of our
Galaxy: a first-overtone pulsator XZ~Ceti (Szabados \etal 2007).
However, thanks to a large number of anomalous Cepheids discovered by the OGLE
survey in the Magellanic Clouds (Soszyñski \etal 2015a) we demonstrated
that coefficients $\phi_{21}$ and $\phi_{31}$ derived from
the Fourier light curve decomposition are useful discriminants between
various types of Cepheids and RR~Lyr stars. Using the Fourier
analysis, Soszyñski \etal (2015a) discovered four Galactic anomalous
Cepheids in the foreground of the Magellanic Clouds -- the first
fundamental-mode anomalous Cepheids known in the Milky Way halo.

\MakeTable{l@{\hspace{3pt}}
c@{\hspace{6pt}}
c@{\hspace{8pt}}
c@{\hspace{8pt}}
c@{\hspace{8pt}}
c@{\hspace{8pt}}
c@{\hspace{3pt}}}
{12.5cm}{Galactic Anomalous Cepheids in the Foreground of the Magellanic Clouds}
{\hline
\noalign{\vskip3pt}
\multicolumn{1}{c}{Identifier}
& Pulsation
& $P$
& $\langle{I}\rangle$
& $\langle{V}\rangle$
& R.A.
& Dec. \\
& mode 
& [d]
& [mag]
& [mag]
& [J2000.0]
& [J2000.0] \\
\noalign{\vskip3pt}
\hline
\noalign{\vskip3pt}
OGLE-GAL-ACEP-001 & F  & 1.3204588 & 14.405 & 14.904 & 23\uph59\upm14\zdot\ups51 & $-68\arcd13\arcm56\zdot\arcs6$ \\
OGLE-GAL-ACEP-002 & F  & 0.8337190 & 16.824 & 17.376 & 01\uph10\upm57\zdot\ups51 & $-71\arcd01\arcm57\zdot\arcs6$ \\
OGLE-GAL-ACEP-003 & F  & 1.0701226 & 14.927 & 15.445 & 01\uph20\upm14\zdot\ups74 & $-65\arcd42\arcm38\zdot\arcs3$ \\
OGLE-GAL-ACEP-004 & F  & 0.7978865 & 15.520 & 16.185 & 02\uph56\upm08\zdot\ups54 & $-79\arcd38\arcm08\zdot\arcs5$ \\
OGLE-GAL-ACEP-005 & 1O & 0.5041628 & 15.719 & 16.159 & 04\uph22\upm11\zdot\ups04 & $-66\arcd48\arcm08\zdot\arcs7$ \\
OGLE-GAL-ACEP-006 & F  & 1.8836187 & 12.461 & 13.014 & 04\uph23\upm51\zdot\ups47 & $-76\arcd54\arcm42\zdot\arcs8$ \\
OGLE-GAL-ACEP-007 & 1O & 0.5201967 & 15.138 & 15.933 & 08\uph10\upm34\zdot\ups29 & $-70\arcd02\arcm34\zdot\arcs8$ \\
\noalign{\vskip3pt}
\hline}

In this study, we extend our sample of Galactic anomalous Cepheids to
seven: five fundamental-mode and two first-overtone pulsators. All these
stars have light curve shapes typical for anomalous Cepheids, but they are
much brighter than their counterparts in the Magellanic System. Thus, we
assume that they are Galactic variables. Table~2 gives their basic
properties, while Fig.~3 shows their positions in the sky.

Note that the OCVS also includes Galactic RR~Lyr stars located in the
foreground of the Magellanic Clouds. However, it is impossible to
completely separate RR~Lyr stars belonging to the Milky Way, LMC, and SMC,
because halos of these three galaxies overlap with each other. For this
reason the Galactic RR~Lyr stars have designations which follow the scheme
for Magellanic Cloud variables (OGLE-LMC-RRLYR-NNNNNN or
OGLE-SMC-RRLYR-NNNNNN). For anomalous Cepheids the situation is different
-- for our seven objects there is no doubt that all of them are members of
the Milky Way. Therefore we propose a new designation for Galactic
anomalous Cepheids: OGLE-GAL-ACEP-NNN, where NNN is a three-digit
consecutive number. Their {\it I}- and {\it V}-band OGLE light curves can
be downloaded from the FTP site
\begin{center}
{\it ftp://ftp.astrouw.edu.pl/ogle/ogle4/OCVS/gal/acep/}\\
\end{center}

Our investigation shows that anomalous Cepheids must be quite numerous in
the Milky Way halo. We found on average one Galactic anomalous Cepheid per
100 square degrees of the sky, so simple scaling to the whole sphere gives
more than 400 such pulsators in our Galaxy. We expect that a number of
anomalous Cepheids will be identified in the OGLE fields in the Galactic
bulge and disk.

\Section{Completeness of the Sample}
Soszyñski \etal (2015b, 2016a) estimated that the completeness of the OCVS
in the central regions of the Magellanic Clouds, which were intensively
monitored during the OGLE-II and OGLE-III projects, is nearly 100\%,
because these regions were independently searched for variable stars at
least three times. In turn, our samples in the outer fields were affected
by the gaps between CCD detectors of the mosaic camera, which effectively
reduced the completeness by about 7\%. The main goal of this investigation
was to fill these gaps.

Our new sample of 115 classical Cepheids increases by 6.9\% the total
sample of Cepheids in the outer regions of the Magellanic Clouds (outside
the OGLE-III fields). This is virtually equal to the expected 7\% of the
missed objects in the previous edition of the OCVS. Moreover, the OGLE-IV
fields seem to cover the entire area where classical Cepheids in the
Magellanic System are present, thus we are convinced that the OGLE
Collection of Classical Cepheids in the Magellanic System is now
practically complete. Thus, with this final update of the OCVS we conclude
the work started by Henrietta Leavitt (1908) over a century ago.

Of course, it is still possible that a few additional classical Cepheids
can be hidden in the far outskirts of the Magellanic Clouds, outside the
OGLE fields. We could also miss some objects located very close to bright
stars overexposed on the OGLE reference images and masked around during the
photometric reduction processes (\eg Udalski \etal 2016). Other Cepheids
that can be overlooked are pulsators with unusual behavior, for example
single-mode second-overtone Cepheids which show very small amplitudes and
sinusoidal light curves. Nevertheless, the number of such objects is
certainly negligible.

In turn, the new sample of RR~Lyr stars enlarges the OGLE collection of
these variables in the outer fields by 5.5\%. It is more difficult to
detect RR~Lyr variables than Cepheids, because they are fainter and many of
them show the Blazhko effect which adds noise to the phased light curves.
Therefore the completeness of RR~Lyr stars must be lower, but it is still
above 95\%. Obviously, this value refers to the area of the sky covered by
the OGLE fields. As one can see in Fig.~2, the old stellar population in
the Magellanic System extends in every direction from the LMC and SMC
centers and reaches regions beyond the OGLE-IV sky coverage. The same
remark applies to anomalous Cepheids, which show similar spatial
distribution to RR~Lyr stars (Soszyñski \etal 2015a).

We cross-matched our updated collection with recently published lists of
pulsating stars in the Magellanic Clouds. Kim \etal (2014) identified
4\,212 candidates for Cepheids of various types and 26\,855 candidates for
RR~Lyr stars in the light curve database collected by the EROS-2
microlensing survey. Our sample includes in total 26\,685 Cepheids and
RR~Lyr objects from these lists, although in some cases our detailed
classification does not match the Kim's \etal (2014) suggestions. The
remaining over 4000 candidates for classical pulsating stars published by
Kim \etal (2014) seem to be variable stars of other types.

The OGLE extended photometric database contains {\it I}-band light curves
of 298 out of 299 candidates for Cepheids in the SMC detected by Moretti
\etal (2016) based on the near-infrared VMC survey. We confirm that 39 of
these stars are real classical Cepheids and two objects are anomalous
Cepheids. Most of the remaining 257 VMC candidates for Cepheids are
constant or nearly constant stars.

The sample of 599 classical, anomalous and type~II Cepheids and 2595 RR~Lyr
stars identified by the Gaia Consortium in the outskirts of the LMC
(Clementini \etal 2016) was discussed in detail by Udalski \etal (2016).
The current version of the OCVS contains 2845 out of 3194 Gaia candidates
for Cepheids and RR~Lyr stars, although in several cases our classification
differs from that proposed by Clementini \etal (2016). For example three
Gaia Cepheids are re-classified as RR~Lyr variables. The missing objects
were also analyzed by Udalski \etal (2016) -- most of them lie outside the
OGLE-IV fields, but the Gaia sample includes also a few dozen misclassified
objects.

\Section{Summary}
We presented here an updated version of the OGLE Collection of Cepheids and
RR~Lyr Stars in the Magellanic System. A collective effort of many
astronomers started by Henrietta Leavitt (1908) finally led to the
discovery of practically all classical Cepheids in the Magellanic System
and of the vast majority of other types of classical pulsating stars. The
OCVS offers high-quality, long-term light curves obtained in the standard
{\it I}- and {\it V}-band filters, well suited for studying properties of
pulsating stars and their host galaxies.

Studies of exotic pulsation modes, period changes, phase and amplitude
modulations, mode switching, mapping of the three-dimensional distribution
of the young and old stellar population in the Magellanic System, exploring
the star formation history and past interactions between the Clouds and
Milky Way, mapping of the interstellar extinction, calibration of the
period--luminosity relations, tracing metallicity gradients are another
examples of investigations that can be conducted with the OGLE sample of
Cepheids and RR~Lyr stars.

\Acknow{We would like to thank Profs. M.~Kubiak and G.~Pietrzyñ\-ski,
  former members of the OGLE team, for their contribution to the collection
  of the OGLE photometric data over the past years. We are grateful to
  Z.~Ko³aczkowski and A.~Schwar\-zen\-berg-Czerny for providing software
  used in this study.

  This work has been supported by the National Science Centre, Poland,
  grant MAESTRO 2016/22/A/ST9/00009. The OGLE project has received
  funding from the Polish National Science Centre grant MAESTRO no.
  2014/14/A/ST9/00121.}

\end{document}